\documentstyle [12pt] {article}

\catcode`@=12
\begin{document}
\thispagestyle{empty}
\begin{flushright}
UCRHEP-T111\\
June 1993\
\end{flushright}
\vspace{0.5in}
\begin{center}
{\large \bf Supersymmetric SU(3) $\times$ U(1) Gauge Model:\\
Higgs Structure at the Electroweak Energy Scale\\}
\vspace{0.3in}
\vspace{1.5in}
{\bf T. V. Duong and Ernest Ma\\}
\vspace{0.3in}
{\sl Department of Physics\\}
{\sl University of California\\}
{\sl Riverside, California 92521\\}
\vspace{1.5in}
\end{center}
\begin{abstract}\
We consider a supersymmetric version of the recently proposed SU(3) $\times$
U(1) extended gauge model.  We show that it is possible to have only two Higgs
doublets at the SU(2) $\times$ U(1) energy scale but they are not those of
the minimal supersymmetric standard model.  In particular, the upper bound on
the lightest scalar boson of this model is $4M_Z \sin \theta_W$ at tree
level and goes up to about 189 GeV after radiative corrections.
\end{abstract}

\newpage
\baselineskip 24pt

The quartic scalar couplings of a given supersymmetric gauge theory are
determined by the gauge couplings as well as other possible couplings
appearing in the superpotential.  In the minimal supersymmetric standard
model (MSSM), there are only two Higgs-doublet superfields from which a
cubic invariant cannot be formed.  Hence the quartic scalar couplings of
the Higgs potential $V$ depend only on $g_1$ and $g_2$, {\it i.e.} the
U(1) and SU(2) gauge couplings respectively.  Let
\begin{eqnarray}
V &=& \mu_1^2 \Phi_1^\dagger \Phi_1 + \mu_2^2 \Phi_2^\dagger \Phi_2 +
\mu_{12}^2 (\Phi_1^\dagger \Phi_2 + \Phi_2^\dagger \Phi_1) \nonumber \\
&+& {1 \over 2} \lambda_1 (\Phi_1^\dagger \Phi_1)^2 + {1 \over 2} \lambda_2
(\Phi_2^\dagger \Phi_2)^2 + \lambda_3 (\Phi_1^\dagger \Phi_1) (\Phi_2^\dagger
\Phi_2) \nonumber \\ &+& \lambda_4 (\Phi_1^\dagger \Phi_2) (\Phi_2^\dagger
\Phi_1) + {1 \over 2} \lambda_5 (\Phi_1^\dagger \Phi_2)^2 + {1 \over 2}
\lambda_5^* (\Phi_2^\dagger \Phi_1)^2,
\end{eqnarray}
then in the MSSM,
\begin{equation}
\lambda_1 = \lambda_2 = {1 \over 4} (g_1^2 + g_2^2), ~~ \lambda_3 =
- {1 \over 4} g_1^2 + {1 \over 4} g_2^2, ~~ \lambda_4 = - {1 \over 2} g_2^2,
{}~~ \lambda_5 = 0.
\end{equation}
However, if the standard model is really the remnant\cite{1} of a larger
theory such as SU(3) $\times$ U(1) as proposed recently\cite{2,3} and it is
supersymmetric, then at the electroweak energy scale, after the heavier
particles have been integrated out, the reduced Higgs sector may contain
only two doublets but Eq. (2) is no longer valid.  The reason is the
appearance of possible cubic invariants in the superpotential according
to the larger theory which have no analog in the MSSM.  A first example
based on an $\rm E_6$-inspired left-right model has already been
discovered.\cite{4}

In this paper, we propose a supersymmetric version of SU(3) $\times$ U(1)
which is interesting in its own right.  We discuss how quark and lepton
masses can be generated.  We then focus on a specific scenario where only
two Higgs doublets are relevant at the SU(2) $\times$ U(1) scale.
In particular, we show that the upper bound on the lightest scalar boson
of this model is $4M_Z \sin \theta_W$ at tree level and goes up to about
189 GeV after radiative corrections.  This bound is substantially higher
than the 115 GeV of the MSSM or the 120 GeV of the left-right model
mentioned above.\cite{4}

The salient feature of the new SU(3) $\times$ U(1) model\cite{2,3} is in the
choice of the electric-charge operator within SU(3).  Instead of the usual
$Q = I_3 + Y/2$, it is assumed here that $Q = I_3 + 3Y/2$.  Hence for
SU(3) $\times$ U(1), we have
\begin{equation}
Q = I_3 + {3 \over 2} Y + Y',
\end{equation}
where $Y'$ is the U(1) hypercharge.  Consider now the fermionic content of
this model.  The three families of leptons transform identically as
(3$^*$,0).  Specifically, $(\ell^c, \nu_\ell, \ell)_L$ form an antitriplet
with $I_3 = (0, 1/2, -1/2)$ and $Y = (2/3, -1/3, -1/3)$.  The quarks are
different: the third family $(T,t,b)_L$ is also an antitriplet (3$^*$, 2/3),
but the first two, $(u,d,D)_L$ and $(c,s,S)_L$, are triplets (3, $-1/3$)
with $I_3 = (1/2, -1/2, 0)$ and $Y = (1/3, 1/3, -2/3)$.  All the
charge-conjugate quark states are singlets.  As shown in Refs. [2] and [3],
this structure ensures the absence of all axial-vector anomalies.

The Higgs sector of this model must consist of at least three complex triplets
$(\eta^+, \eta^0, \eta^-)$, $(\rho^0, \rho^-, \rho^{--})$, and $(\chi^{++},
\chi^+, \chi^0)$, transforming as (3,0), (3,$-1$), and (3,1) respectively.
At the first step of symmetry breaking, $\chi^0$ acquires a large vacuum
expectation value, so that SU(3) $\times$ U(1) breaks down to the standard
SU(2) $\times$ U(1) and the exotic quarks $D$, $S$ (of electric charge $-4/3$)
and $T$ (of electric charge 5/3) become massive.  The subsequent breaking of
SU(2) $\times$ U(1) is accomplished with nonzero values of $\langle \eta^0
\rangle$ and $\langle \rho^0 \rangle$, such that $t$, $s$, and $d$ acquire
masses proportional to the former and $b$, $c$, and $u$ acquire masses
proportional to the latter.  To obtain lepton masses, a Higgs sextet was
proposed.\cite{3,5}  However, we would like to adopt a simple alternative.
Let $E_L$ and $E_L^c$ be singlets (1,$-1$) and (1,1), then $E_L E_L^c$ is
an allowed mass term and the mass matrix linking $(\ell_L, E_L)$ to
$(\ell_L^c, E_L^c)$ is of the see-saw form with $\langle \chi^0 \rangle$
contributing to $E_L \ell_L^c$ and $\langle \rho^0 \rangle$ to $\ell_L
E_L^c$ respectively.

We now impose supersymmetry.  In addition to changing all fields into
superfields, we need to add three complex scalar superfields $(\eta'^+,
\eta'^0, \eta'^-)$, $(\rho'^{++}, \rho'^+, \rho'^0)$, and $(\chi'^0, \chi'^-,
\chi'^{--})$, transforming as ($3^*$,0), ($3^*$,1), and $(3^*,-1)$
respectively.  These are required for the cancellation of anomalies
generated by the $\rho$, $\eta$, and $\chi$ superfields.  They also have
invariant couplings to the quark superfields, so that $m_T$ comes from
$\langle \chi^0 \rangle$, but $m_S$ and $m_D$ come from $\langle \chi'^0
\rangle$; $m_t$ comes from $\langle \eta^0 \rangle$, but $m_c$ and $m_u$
come from $\langle \rho'^0 \rangle$; $m_b$ comes from $\langle \rho^0
\rangle$, but $m_s$ and $m_d$ come from $\langle \eta'^0 \rangle$.
Furthermore, the superpotential now contains two cubic invariants
$f \epsilon_{ijk} \eta_i \rho_j \chi_k$ and $f' \epsilon_{ijk} \eta'_i
\rho'_j \chi'_k$ which contribute to the Higgs potential.

The Higgs sector of our supersymmetric version of the SU(3) $\times$ U(1)
model now has 3 triplets and 3 antitriplets.  The part of the Higgs
potential related to the gauge interactions through supersymmetry is
given by
\begin{eqnarray}
V_D &=& {1 \over 2} G_1^2 [ - \rho_i^* \rho_i + \chi_i^* \chi_i + \rho_i'^*
\rho_i' - \chi_i'^* \chi_i' ]^2 \nonumber \\ &+& {1 \over 8} G_3^2 \sum_a
[ \eta_i^* \lambda_{ij}^a \eta_j + \rho_i^* \lambda_{ij}^a \rho_j + \chi_i^*
\lambda_{ij}^a \chi_j - \eta_i'^* \lambda_{ij}^{*a} \eta_j' - \rho_i'^*
\lambda_{ij}^{*a} \rho_j' - \chi_i'^* \lambda_{ij}^{*a} \chi_j' ]^2,
\end{eqnarray}
where $G_1$ and $G_3$ are the U(1) and SU(3) gauge couplings respectively and
$\lambda_{ij}^a$ are the 8 conventional 3 $\times$ 3 SU(3) representation
matrices.  Similarly, the part of the Higgs potential related to the
superpotential is given by
\begin{eqnarray}
V_F &=& f^2 \sum_k [ |\epsilon_{ijk} \eta_i \rho_j |^2 + |\epsilon_{ijk}
\rho_i \chi_j |^2 + |\epsilon_{ijk} \chi_i \eta_j |^2 ] \nonumber \\
&+& f'^2 \sum_k [ |\epsilon_{ijk} \eta_i' \rho_j' |^2 + |\epsilon_{ijk}
\rho_i' \chi_j' |^2 + |\epsilon_{ijk} \chi_i' \eta_j' |^2 ].
\end{eqnarray}
Let $\langle \chi^0 \rangle = u \neq 0$ and $\langle \chi'^0 \rangle = u'
\neq 0$, then the SU(3) $\times$ U(1) gauge symmetry is broken down to the
standard SU(2) $\times$ U(1).  Five of the twelve degrees of freedom
contained in $\chi$ and $\chi'$ are absorbed into the five vector gauge
bosons which become massive.  The remaining seven are heavy physical states.
As for $\eta, \rho, \eta'$, and $\rho'$, if their doublet components are to
be light, then their singlet components are necessarily heavy because their
mass terms depend differently on $u^2$ and $u'^2$.  In general, the reduced
Higgs potential will contain four doublets if $\langle \eta^0 \rangle,
\langle \rho^0 \rangle, \langle \eta'^0 \rangle$, and $\langle \rho'^0
\rangle$ are all nonzero.  However, an interesting alternative exists if
we assume an extra discrete $Z_2$ symmetry under which $u_L^c, d_L^c, s_L^c$,
and $c_L^c$ are odd and all other superfields are even.  It has been
shown\cite{6} that the breaking of this $Z_2$ by soft terms which also
break the supersymmetry will allow $u, d, s$, and $c$ to acquire radiative
masses in one-loop order through gluino exchange.  Hence it is possible
for $\langle \eta'^0 \rangle$ and $\langle \rho'^0 \rangle$ to be zero
so that $\eta'$ and $\rho'$ may be assumed heavy and will not appear in the
reduced Higgs potential at the electroweak energy scale.

Redefine $(- \overline {\rho^-}, \overline {\rho^0})$ as $\Phi_1$ and
$(\eta^+, \eta^0)$ as $\Phi_2$, then the parts of $V_D$ and $V_F$ which
contain $\Phi_1, \Phi_2, \chi^0$, and $\chi'^0$ are given by
\begin{eqnarray}
V' &=& {1 \over 2} G_1^2 [(\Phi_1^\dagger \Phi_1)^2 - 2 (\Phi_1^\dagger
\Phi_1) (|\chi^0|^2 - |\chi'^0|^2) + (|\chi^0|^2 - |\chi'^0|^2)^2] \nonumber
\\ &+& {1 \over 6} G_3^2 [(\Phi_1^\dagger \Phi_1 + \Phi_2^\dagger \Phi_2)^2
- 3 (\Phi_1^\dagger \Phi_2) (\Phi_2^\dagger \Phi_1) \nonumber \\ &~& ~~~~~~
- (\Phi_1^\dagger \Phi_1
+ \Phi_2^\dagger \Phi_2) (|\chi^0|^2 - |\chi'^0|^2) + (|\chi^0|^2 -
|\chi'^)|^2)^2] \nonumber \\ &+& f^2 [(\Phi_1^\dagger \Phi_2) (\Phi_2^\dagger
\Phi_1) + (\Phi_1^\dagger \Phi_1 + \Phi_2^\dagger \Phi_2) |\chi^0|^2].
\end{eqnarray}
Since $\langle \chi^0 \rangle = u$ and $\langle \chi'^0 \rangle = u'$, there
are cubic interactions in $V'$ involving $\chi^0$ and $\Phi_{1,2}$ as well
as $\chi'^0$ and $\Phi_{1,2}$.  These have to be taken into account\cite{1,4}
in obtaining the effective quartic scalar couplings $\lambda_i$ of Eq. (1).
However, because $\sqrt 2 Re \chi^0$ and $\sqrt 2 Re \chi'^0$ are not mass
eigenstates, we need to consider their 2 $\times$ 2 mass-squared matrix
given by
\vglue 0.1cm
\begin{equation}
{\cal M}^2 = \left( \begin{array} {c@{\quad}c} M^2 \cos^2 \gamma + M'^2
\sin^2 \gamma & - (M^2 + M'^2) \sin \gamma \cos \gamma \\ - (M^2 + M'^2)
\sin \gamma \cos \gamma & M^2 \sin^2 \gamma + M'^2 \cos^2 \gamma \end{array}
\right),
\end{equation}
\vglue 0.2cm
\noindent where $M^2 = 2 (G_1^2 + G_3^2/3) (u^2 + u'^2)$, $\tan \gamma \equiv
u'/u$,
and $M'$ is the mass of the heavy pseudoscalar boson $\sqrt 2 (\sin \gamma
Im \chi^0 - \cos \gamma Im \chi'^0)$ which has no cubic coupling to
$\Phi_{1,2}$.  The determinant of ${\cal M}^2$ is equal to $M^2 M'^2 \cos^2
2 \gamma$.  Hence
\begin{eqnarray}
\lambda_1 &=& {1 \over 3} G_3^2 + G_1^2 - {{2(u^2 + u'^2)} \over {M^2 M'^2
\cos^2 2 \gamma}} [ (f^2 - G_1^2 - G_3^2/6)^2 \cos^2 \gamma ({\cal M}^2)_{22}
\nonumber \\ &~& ~~~~~~~~ - 2 (f^2 - G_1^2 - G_3^2/6)(G_1^2 + G_3^2/6)
\sin \gamma \cos \gamma ({\cal M}^2)_{12} \nonumber \\ &~& ~~~~~~~~ +
(G_1^2 + G_3^2/6)^2 \sin^2 \gamma ({\cal M}^2)_{11} ], \\
\lambda_2 &=& {1 \over 3} G_3^2 - {{2(u^2+u'^2)} \over {M^2 M'^2
\cos^2 2 \gamma}} [(f^2 - G_3^2/6)^2 \cos^2 \gamma ({\cal M}^2)_{22}
\nonumber \\ &~& ~~~ - 2 (f^2 - G_3^2/6)(G_3^2/6) \sin \gamma
\cos \gamma ({\cal M}^2)_{12} + (G_3^2/6)^2 \sin^2 \gamma
({\cal M}^2)_{11}], \\
\lambda_3 &=& {1 \over 3} G_3^2 - {{2(u^2+u'^2)} \over {M^2 M'^2 \cos^2
2 \gamma}} [(f^2-G_3^2/6)(f^2 - G_1^2 -G_3^2/6) \cos^2 \gamma ({\cal M}^2)_{22}
\nonumber \\ &~& ~~~ - [(f^2 - G_3^2/6)(G_1^2 + G_3^2/6) + (f^2 - G_1^2 -
G_3^2/6)
(G_3^2/6)] \sin \gamma \cos \gamma ({\cal M}^2)_{12} \nonumber \\
&~& ~~~ +(G_3^2/6)(G_1^2 +
G_3^2/6) \sin^2 \gamma ({\cal M}^2)_{11} ], \\
\lambda_4 &=& - {1 \over 2} G_3^2 + f^2, ~~~ \lambda_5 = 0.
\end{eqnarray}
In the limit $f=0$,
\begin{equation}
\lambda_1 = \lambda_2 = {{G_3^2 (G_3^2 + 4 G_1^2)} \over {4 (G_3^2 +
3 G_1^2)}}, ~~~ \lambda_3 = {{G_3^2 (G_3^2 + 2 G_1^2)} \over {4 (G_3^2 +
3 G_1^2)}}, ~~~ \lambda_4 = - {1 \over 2} G_3^2, ~~~ \lambda_5 = 0.
\end{equation}
Assuming the tree-level relations $g_2 = G_3$ and $g_1^{-2} = G_1^{-2} +
3 G_3^{-2}$, we then have $G_1^2 = g_1^2 g_2^2 / (g_2^2 - 3 g_1^2)$ and
the MSSM conditions, {\it i.e.} Eq. (2), are obtained as expected.\cite{4}

Since $f \neq 0$ in the general case, the Higgs potential of this model
differs from that of the MSSM even though there are only two Higgs doublets
at the electroweak energy scale.  It also differs from that of the left-right
model mentioned previously.\cite{4}  The $f^2$ and $f^4$ terms in
$\lambda_{1,2,3}$ depend on $\gamma$ and the $f^4$ terms on $M^2/M'^2$ as
well.  For illustration, let us take the special case $\cos \gamma = 1$, then
\begin{eqnarray}
\lambda_1 &=& {1 \over 4} (g_1^2 + g_2^2) + f^2 \left( 1 + {{3g_1^2} \over
g_2^2} \right) - {{3f^4} \over g_2^2} \left( 1 - {{3g_1^2} \over g_2^2}
\right), \\
\lambda_2 &=& {1 \over 4} (g_1^2 + g_2^2) + f^2 \left( 1 - {{3g_1^2} \over
g_2^2} \right) - {{3f^4} \over g_2^2} \left( 1 - {{3g_1^2} \over g_2^2}
\right), \\
\lambda_3 &=& - {1 \over 4} g_1^2 + {1 \over 4} g_2^2 + f^2 - {{3f^4} \over
g_2^2} \left( 1 - {{3g_1^2} \over g_2^2} \right), \\
\lambda_4 &=& - {1 \over 2} g_2^2 + f^2, ~~~ \lambda_5 = 0.
\end{eqnarray}
The requirement that $V$ be bounded from below puts an upper bound on $f^2$
so that
\begin{equation}
0 \leq f^2 \leq {1 \over 2} g_2^2.
\end{equation}
Let us now specialize further to the case $f = f_{max}$, we then find
\begin{equation}
\lambda_1 = 4 g_1^2, ~~~ \lambda_2 = g_1^2, ~~~ \lambda_3 = 2 g_1^2, ~~~
\lambda_4 = \lambda_5 = 0.
\end{equation}
The equality of $\lambda_4$ and $\lambda_5$ means that an accidental custodial
SU(2) symmetry exists\cite{1} so that the charged Higgs boson $H^\pm$ and the
pseudoscalar Higgs boson $A$ form a triplet with a common mass given by
\begin{equation}
m_A^2 = {{-2\mu_{12}^2} \over {\sin 2 \beta}},
\end{equation}
where $\tan \beta \equiv \langle \phi_2^0 \rangle / \langle \phi_1^0 \rangle$.
The 2 $\times$ 2 mass-squared matrix spanning $\sqrt 2 Re \phi_1^0$ and
$\sqrt 2 Re \phi_2^0$ is now
\begin{equation}
{\cal M}^2 = \left( \begin{array} {c@{\quad}c} 16 M_Z^2 \sin^2 \theta_W
\cos^2 \beta + m_A^2 \sin^2 \beta & (8 M_Z^2 \sin^2 \theta_W - m_A^2)
\sin \beta \cos \beta \\ (8 M_Z^2 \sin^2 \theta_W - m_A^2) \sin \beta
\cos \beta & 4 M_Z^2 \sin^2 \theta_W \sin^2 \beta + m_A^2 \cos^2 \beta +
\epsilon / \sin^2 \beta \end{array} \right),
\end{equation}
where
\begin{equation}
\epsilon = {{3g_2^2 m_t^2} \over {8 \pi^2 M_W^2}} \ln \left( 1 + {{\tilde
m^2} \over m_t^2} \right)
\end{equation}
comes from radiative corrections due to the $t$ quark and its two scalar
supersymmetric partners of effective mass $\tilde m$.  This implies
\begin{equation}
m_h^2 \leq 4 M_Z^2 \sin^2 \theta_W (1 + \cos^2 \beta)^2 + \epsilon
\end{equation}
as well as
\begin{equation}
m_h^2 \leq {{m_A^2 (1 + \cos^2 \beta)^2 + 4 \epsilon \cot^2 \beta} \over
{1 + 3 \cos^2 \beta}},
\end{equation}
where $h$ is the lighter of the two mass eigenstates.  Hence $m_h$ has an
upper bound of $4 M_Z \sin \theta_W$ at tree level and it goes up to about
189 GeV after radiative corrections assuming $m_t$ = 150 GeV and $\tilde m$
= 1 TeV.

In conclusion, we have presented in this paper a supersymmetric SU(3) $\times$
U(1) model which has a possible reduction to the standard SU(2) $\times$ U(1)
model with two Higgs doublets at the electroweak energy scale.  Because of
the existence of cubic invariants in the superpotential of the larger
theory, the reduced Higgs potential is not that of the minimal supersymmetric
standard model (MSSM).  The quartic scalar couplings are given by Eqs. (8) to
(11), instead of Eq. (2).  For illustration, we have taken the special case
$\cos \gamma = 1$ ({\it i.e.} neglecting $\langle \chi'^0 \rangle$) and then
$f = g_2/\sqrt 2$ ({\it i.e.} $f = f_{max}$), resulting in the very simple
conditions of Eq. (18).  We then show that instead of 115 GeV in the MSSM
or 120 GeV in a left-right model discussed elsewhere\cite{4}, the upper
bound of the lightest scalar boson in this model is $4 M_Z \sin \theta_W$
and goes up to 189 GeV after radiative corrections assuming $m_t$ = 150 GeV
and $\tilde m$ = 1 TeV.  In the future, as it becomes possible experimentally
to explore the Higgs sector at the electroweak energy scale, it is important
to realize that even if supersymmetry exists, the MSSM is not the only
possibility for two Higgs doublets.
\vspace{0.3in}
\begin{center} {ACKNOWLEDGEMENT}
\end{center}

This work was supported in part by the U. S. Department of Energy under
Contract No. DE-AT03-87ER40327.

\newpage
\bibliographystyle{unsrt}

\end{document}